\documentclass[aps,prd,reprint,superscriptaddress,nofootinbib,floatfix,longbibliography]{revtex4-1}
\usepackage{amsmath}
\usepackage{amsfonts}
\usepackage{bm}
\usepackage{graphicx}
\usepackage{hyperref}
\usepackage{soul}
\usepackage{color}
\usepackage{comment}

\enlargethispage{\baselineskip}

\hypersetup{colorlinks   = true,
            urlcolor     = blue,
            citecolor    = blue,
            linkcolor    = blue,
            menucolor    = blue,
            anchorcolor  = blue,
            filecolor    = blue
}
\everymath{\displaystyle}

\begin{document}

\title{Astrophysical neutrino oscillations after pulsar timing array analyses}

\author{Gaetano Lambiase}
\email{lambiase@sa.infn.it}
\affiliation{Dipartimento di Fisica ``E.R.\ Caianiello'', Universit\`a degli Studi di Salerno,\\ Via Giovanni Paolo II, 132 - 84084 Fisciano (SA), Italy}
\affiliation{Istituto Nazionale di Fisica Nucleare - Gruppo Collegato di Salerno - Sezione di Napoli,\\ Via Giovanni Paolo II, 132 - 84084 Fisciano (SA), Italy}

\author{Leonardo Mastrototaro}
\email{lmastrototaro@unisa.it}
\affiliation{Dipartimento di Fisica ``E.R.\ Caianiello'', Universit\`a degli Studi di Salerno,\\ Via Giovanni Paolo II, 132 - 84084 Fisciano (SA), Italy}
\affiliation{Istituto Nazionale di Fisica Nucleare - Gruppo Collegato di Salerno - Sezione di Napoli,\\ Via Giovanni Paolo II, 132 - 84084 Fisciano (SA), Italy}

\author{Luca Visinelli}
\email{luca.visinelli@sjtu.edu.cn\\}
\affiliation{Tsung-Dao Lee Institute (TDLI), 520 Shengrong Road, 201210 Shanghai, P.\ R.\ China}
\affiliation{School of Physics and Astronomy, Shanghai Jiao Tong University, 800 Dongchuan Road, 200240 Shanghai, P.\ R.\ China}

\begin{abstract}
The pattern of neutrino flavor oscillations could be altered by the influence of noisy perturbations such as those arising from a gravitational wave background (GWB). A stochastic process that is consistent with a GWB has been recently reported by the independent analyses of pulsar timing array (PTA) data sets collected over a decadal timescale by the North American Nanohertz Observatory for Gravitational Waves, the European Pulsar Timing Array jointly with the Indian Pulsar Timing Array, the Parkes Pulsar Timing Array, and the Chinese Pulsar Timing Array collaborations. We investigate the modifications in the neutrino flavor oscillations under the influence of the GWB reported by the PTA collaborations and we discuss how such effects could be potentially revealed in near-future neutrino detectors, possibly helping the discrimination of different models for the GWB below the nHz frequency range.
\end{abstract}

\date{\today}
\maketitle

\section{Introduction}
\label{sec:introduction}

Neutrinos are ideal astrophysical messengers owing to their distinctive properties such as feeble interactions and neutrality, which allow them to reach us from the cosmic accelerator where they had originated avoiding absorption and deflection by magnetic fields~\cite{Greisen:1960wc, Reines:1960we}. The diffuse neutrino emission as well as unresolved point sources within our Galaxy are being observed in various channels at present and near-future detectors~\cite{doi:10.1126/science.adc9818}.

One peculiar neutrino feature is flavor oscillations, a purely quantum mechanical phenomenon related to the mass split of the three mass eigenvalues~\cite{Gribov:1968kq, Bilenky:1978nj}, confirmed at underground detectors collecting neutrino fluxes both from baseline facilities~\cite{OPERA:2010pne, Vogel:2015wua} and of extraterrestrial origin~\cite{Super-Kamiokande:1998kpq, Super-Kamiokande:2001ljr, IceCube:2017lak}. Neutrino oscillations transform the flavor composition according to the Pontecorvo–Maki–Nakagawa–Sakata matrix~\cite{Learned:1994wg}, which accounts for the mismatch between the interaction and mass bases of these particles. The neutrino flavor evolution can be determined using the global best-fit mixing parameters~\cite{Gonzalez-Garcia:2014bfa}, so that the observation of a ratio inconsistent with the expected flux would be a signal of new physics in the neutrino sector provided a precise source production model and a good spatial sensitivity of the detector. Indeed, the propagation of neutrinos could be also affected by other phenomena such as neutrino decay~\cite{Beacom:2002vi, Baerwald:2012kc}, the existence of sterile neutrinos~\cite{Athar:2000yw}, pseudo-Dirac neutrinos~\cite{Beacom:2003eu, Esmaili:2009fk}, dark matter~\cite{deSalas:2016svi}, Lorentz or CPT violation~\cite{Hooper:2005jp}, and quantum gravity-induced decoherence~\cite{Anchordoqui:2005gj}.

A different framework in which neutrino oscillations would help investigate new physics involves the stochastic fluctuations induced by a noisy background~\cite{Loreti:1994ry}, such as what is provided by the presence of gravitational waves (GWs) released from different mechanisms in galaxies and from the early Universe. The collective signals from all incoherent GW sources in the Universe lead to a background of GWs (GWB), whose strength differs at various frequency windows. Understanding the GWB strength at different wavelengths is crucial to test various cosmological theories as well as the (astro)physics of compact objects, so that it is a primary target for the next generation of GW detectors, including interferometers~\cite{Meacher:2015iua, Abbott:2016cjt} as well as the search at frequencies of the order of the nHz from pulsar timing array (PTA)~\cite{Stinebring:1990px, 1990ApJ...361..300F} or the nature of neutrinos~\cite{King:2023cgv}.

A stochastic source such as the GWB can significantly modify the propagation and flavor oscillations of neutrino packets, leading to observable effects over the coherence length scale and the probability to detect a given flavor~\cite{Ahluwalia:1996ev, Fornengo:1996ef, Cardall:1996cd, Capozziello:1999qm, Lambiase:2004qk, Lambiase:2005gt, Cuesta:2008te, Lambiase:2013haa, Chakraborty:2013ywa, Visinelli:2014xsa, Chakraborty:2015vla, Dvornikov:2019fhi, Koutsoumbas:2019fkn, Dvornikov:2006ji, Buoninfante:2019der, Lambiase:2021txu, Dvornikov:2021hps, Swami:2022xet, Rosado:2011kv, Giudice:2016zpa, Jung:2017flg, Lai:2018rto, Christian:2018vsi, Dai:2018enj, Jung:2018kde, Choi:2018axi, Lambiase:2022ucu}. The precise measurement of the properties of a neutrino wave packet of astrophysical origin could then provide a complementary tool to test GWB models.

In the nHz range, the GWB is actively searched through correlations in the times-of-arrival pulses emitted by an ensemble of pulsars through PTA techniques by several collaborations, including the North American Nanohertz Observatory for Gravitational Waves (NANOGrav) collaboration~\cite{McLaughlin:2013ira}, the European Pulsar Timing Array (EPTA) jointly with the Indian Pulsar Timing Array (InPTA)~\cite{Chen:2021rqp}, the Parkes Pulsar Timing Array (PPTA)~\cite{Goncharov:2021oub}, and the Chinese Pulsar Timing Array (CPTA)~\cite{2016ASPC..502...19L}. Recently, all these collaborations have independently reported the evidence for an excess red common-spectrum signal from data analysis with PTA, that is consistent with the stochastic process of a GWB~\cite{NANOGrav:2023gor, NANOGrav:2023icp, NANOGrav:2023hfp, NANOGrav:2023ctt, NANOGrav:2023hvm, Zic:2023gta, Reardon:2023zen, Reardon:2023gzh, EPTA:2023sfo, EPTA:2023akd, EPTA:2023fyk, EPTA:2023gyr, EPTA:2023xxk, EuropeanPulsarTimingArray:2023egv, Xu:2023wog}. This establishes a relevant milestone in the GW search that follows after a multi-decadal effort and corroborates previous results following the data analyses by NANOGrav 12.5-year~\cite{NANOGrav:2020bcs}, EPTA~\cite{Chen:2021rqp}, PPTA~\cite{Goncharov:2021oub} and the International PTA~\cite{Antoniadis:2022pcn}. The important addition in the most recent analyses marks the existence of inter-pulsar correlations, a fingerprint for the presence of a correlated stochastic GWB signal that allows for discrimination with respect to other potential sources of correlated signals~\cite{Hellings:1983fr}.

One possible explanation for the signals observed with PTA techniques is the incoherent GW background released from inspiral supermassive black hole binaries (SMBHBs), in which systems of binary black holes with total mass $M_{\rm tot} \in [10^6$-$10^{10}]\,M_\odot$ and possibly even heavier contribute to a broadband signal at the nHz frequency~\cite{Ellis:2023dgf}. Exotic components are also expected to lead to a GWB that is potentially of similar amplitude, including cosmic strings~\cite{Kitajima:2023vre, Ellis:2023tsl}, phase transitions in the early Universe~\cite{Addazi:2023jvg, Athron:2023mer, Bai:2023cqj, Fujikura:2023lkn, Zu:2023olm, Yang:2023qlf}, or inflation~\cite{Vagnozzi:2023lwo}, see also Refs.~\cite{Schneider:2010ks, Kuroyanagi:2018csn}.

In this work, we explore the implication of the recent detection of the GWB stemming from the analyses of the PTA data at the nHz frequency range onto the propagation of neutrino bursts, and we comment on how a future detection could help frame the astrophysical model that generates the GWB, even exploring the frequency region below the nHz which is already motivated by known physics~\cite{Neronov:2020qrl, Moore:2021ibq, Brandenburg:2021tmp, DeRocco:2023qae} and in more exotic models~\cite{Chang:2019mza, Freese:2023fcr}. We set $\hbar = c = 1$ unless otherwise specified.

\section{Neutrino oscillations in the presence of a GWB.}

We consider three Dirac neutrino generations $\nu_\alpha$ with flavors $\alpha = e, \mu, \tau$, related to the mass eigenstates $\nu_i$ with $i =1,2,3$ by a unitary neutrino mixing matrix $U$ as $\nu_\alpha = U_{\alpha i}\nu_i$. We follow the standard convention for the neutrino mixing matrix~\cite{Workman:2022ynf}. Since the orientation and amplitude of the GW strain interacting with the neutrinos are random quantities, we switch to the neutrino density matrix description, which evolves in time as $i\dot\rho = [H_{\rm eff}+H_{\rm int}, \rho]$. Here, $H_{\rm eff} = UH^{({\rm vac})}U^\dag$ and $H_{\rm int} = UH^{(g)}U^\dag$ are the effective and the interaction Hamiltonian, respectively, and
\begin{eqnarray}
    H^{({\rm vac})} &=& \frac{1}{2E}\mathrm{diag}\left(0,\Delta m_{12}^2,\Delta m_{13}^2\right) \,,\label{eq:mass-matrix}\\
    H^{(g)} &=& H^{({\rm vac})} \left(A_+ h_+ + A_\times h_\times\right)\,,\label{eq:interaction}
\end{eqnarray}
where $\Delta m_{ij}^2=m_i^2-m_j^2$ ($i,j = 1,2,3$) is the mass squared difference. In the interaction picture with the density matrix $\rho_{\rm int} \equiv e^{iH_{\rm eff}t}\rho e^{-iH_{\rm eff}t}$, the equation of evolution for the neutrino density matrix is~\cite{Loreti:1994ry, Dvornikov:2019jkc, Dvornikov:2020dst, Dvornikov:2021sac},
\begin{equation}
     \label{evolution}
     \frac{\rm d}{{\rm d}t}\langle\rho_{\rm int}\rangle(t) = -\frac{3}{128}\,g(t)\,[H_{\rm eff},[H_{\rm eff},\langle\rho_{\rm int}(t)\rangle]]\,,
\end{equation}
where $g(t) = \sum_\zeta \int_0^t {\rm d}t' \langle h_\zeta(t')h_\zeta(t)\rangle$ accounts for the average of the incoherent GW contribution and a summation over the polarization $\zeta$, while square brackets denote the average over the GWB parameters. Given the emission probability $P_{\sigma}(0)$ for the flavor $\sigma$ produced by a source at a distance $D$ from Earth with energy $E$, the probability to detect the flavor $\lambda$ at Earth is~\cite{Dvornikov:2021sac}
\begin{eqnarray}
    \label{eq:probability}
    P_{\lambda}(D) &=& \sum_\sigma P_{\sigma}(0)\bigg[\sum_i|U_{\lambda i}|^2|U_{\sigma i}|^2\\ 
    && + 2{\rm\,Re}\sum_{i>j}U_{\lambda i}U^*_{\lambda j}U^*_{\sigma i}U_{\sigma j}e^{-i D / L_{\rm osc}^{ij} }\,e^{-\Gamma^{(ij)}}\bigg]\,,\nonumber
\end{eqnarray}
where the indices $i$, $j$ label the neutrino mass eigenstates and the oscillatory part describes the oscillations of the neutrino flavors with the oscillation length $L_{\rm osc}^{ij} = 2E/\Delta m_{ji}^2$. Oscillations in the neutrino flavors are modulated by the coherence term in Eq.~\eqref{eq:probability},
\begin{equation}
    \Gamma^{(ij)} = \left(\frac{3 \,H_0}{8 \,\pi\, L_{\rm osc}^{ij}}\right)^2 \int_{f_{\rm min}}^{f_{\rm max}}\frac{{\rm d}f}{f^5} \sin^2(\pi f D) \Omega_{\rm GW}(f)\,,
    \label{eq:decoherenceterm}
\end{equation}
where $D$ is the distance of the neutrino source. The coherence term depends on the neutrino energy such that there exist a threshold energy $E_{\rm thr}$ below which oscillations are suppressed. The frequency $f_{\rm min}$ describes the lowest frequency at which the energy loss from GW radiation overcomes other forms of dissipation, which for SMBHBs corresponds to dynamical friction and gravitational slingshot in the earliest stages of the encountering~\cite{Quinlan:1996vp, Sesana:2004sp}.

\section{The GWB in the nHz window}

The GWB impacts the propagation of neutrino oscillations and affects the detection probability according to Eq.~\eqref{evolution}. The analysis performed by the PTA collaborations model the dependence of the GWB strain on the frequency via a power-law spectrum of the form
\begin{equation}
	h_c(f) = A_*\,\left(\frac{f}{f_{\rm yr}}\right)^{\frac{3-\gamma}{2}}\,,
 \label{eq:strain}
\end{equation}
where $f_{\rm yr} = 1{\rm\,yr}^{-1}\approx 31.7\,$nHz is the reference frequency. The expression in Eq.~\eqref{eq:strain} translates into the fractional energy density in GWs at the nHz frequency~\cite{Ellis:2020ena}
\begin{equation}
	\label{eq:GWMBHB}
	\Omega_{\rm GW}(f) = \frac{2 \pi^2}{3 H_0^2} f_{\rm yr}^{2}|A_*|^2\,\left(\frac{f}{f_{\rm yr}}\right)^{5-\gamma}\,.
\end{equation}

Given the GW energy density in Eq.~\eqref{eq:GWMBHB}, the neutrino coherence term in Eq.~\eqref{eq:decoherenceterm} can be obtained analytically as
\begin{equation}
    \label{eq:decoherenceterm1}
    \Gamma^{(ij)} \simeq \frac{3}{64(\gamma - 1)}\left(\frac{|A_*|}{f_{\rm yr}L_{\rm osc}^{ij}}\right)^2 \left(\frac{f_{\rm min}}{f_{\rm yr}}\right)^{1 - \gamma}\,,
\end{equation}
where the distance of the neutrino source does not enter the expression above for the values of the combination $D \gg 1/f_{\rm min}$ considered here. Note, that the expression above does not depend on the Hubble constant since both the neutrino sources and the GW perturbations that affect the neutrino oscillations reside within the galaxy. Since $\gamma$ is significantly different from one in the analyses presented by the PTA collaborations, the denominator in Eq.~\eqref{eq:decoherenceterm1} does not pose a threat.

The frequency $f_{\rm min}$ coincides with the lowest frequency considered in the signal-dominated range of the emitted GWs. In the expression above we have assumed that the GW signal from a SMBHB is released within a range spanning the frequency $f_{\rm min}$ at the beginning of the inspiral phase up to $f_{\rm max}$ when the coalescence occurs. Since the integrand in Eq.~\eqref{eq:decoherenceterm} is dominated by the lowest frequencies in the domain, the shape of the GWB spectrum at frequencies $f \gg f_{\rm min}$ does not affect the results. An estimate based on the inspiral time $T_{\rm insp} = 75\,$Myr gives~\cite{Rosado:2011kv}
\begin{equation}
    \label{eq:fmin}
    f_{\rm min} = \left(\frac{256}{5}\,T_{\rm insp}\right)^{-3/8}\!\frac{\pi}{\left(GM_{\rm tot}\right)^{5/8}} \approx 10^{-9}{\rm\,Hz}\,,
\end{equation}
where the numerical result is valid for a total mass $M_{\rm tot} = 10^{10}\,M_\odot$. One may even consider exotic models that predict lower values of the cutoff frequency in the extremely-low-frequency (ELF) band, $f_{\rm min} \sim 10^{-18}\,$Hz, which can be probed from anisotropies in the cosmic microwave background (CMB)~\cite{Krauss:1992ke, Thorne:1995xs}. The integrated energy density is bound by~\cite{Smith:2006nka, Pagano:2015hma, Caprini:2018mtu}
\begin{equation}
    \int \frac{{\rm d}f}{f}\,\Omega_{\rm GW}(f)\,h^2 \leq 5.6 \times 10^{-6}\Delta N_{\rm eff}\,,
\end{equation}
where $N_{\rm eff}$ is the effective number of neutrinos in the thermal bath and $\Delta N_{\rm eff}$ its excess as detectable in the CMB and big-bang nucleosynthesis~\cite{Planck:2018vyg,ParticleDataGroup:2018ovx}.

Oscillations in the flavors of the neutrinos in the wave packet are suppressed below the threshold energy
\begin{equation}
    \label{eq:thresholdE}
    E_{\rm thr} \simeq \frac{\Delta m_{12}^2}{16}\frac{A_*}{f_{\rm yr}}\,\left(\frac{3}{\gamma - 1}\right)^{1/2} \left(\frac{f_{\rm min}}{f_{\rm yr}}\right)^{(1 - \gamma)/2}\,,
\end{equation}
so that $\Gamma^{(12)} \gtrsim 1$. The lower value of the mass square difference $\Delta m_{12}^2$ allows for the minimum threshold energy. For the GWB spectrum reported in Ref.~\cite{NANOGrav:2023gor} and for $f_{\rm min}$ below the nHz range, the threshold energy lies within the range $E_{\rm thr}=$[100\,keV -- 10\,MeV], which is potentially detectable at various neutrino facilities. For example, the oscillation length for the $1 \leftrightarrow 2$ conversion in a neutrino packet of energy $E_\nu \sim 10\,$MeV is around 50\,km, so that suppression in oscillations for neutrinos of astrophysical origin could be realized for a combination of the GWB parameters that leads to the neutrino coherence term $\Gamma^{(12)} \simeq \mathcal{O}(1)$.

\section{Results for a variable power-law exponent}

We first consider the scenario in which the source for the GWB is not specified and the index $\gamma$ in Eq.~\eqref{eq:strain} is a free parameter in the model. For this model, the NANOGrav collaboration reports the amplitude $A_* = 6.4_{-2.7}^{+4.2}\times 10^{-15}$ for frequencies $f \sim 30\,$nHz and a spectral index $\gamma = 3.2_{-0.6}^{+0.6}$~\cite{NANOGrav:2023gor, NANOGrav:2023icp, NANOGrav:2023hfp, NANOGrav:2023ctt, NANOGrav:2023hvm}. PPTA consistently reports the amplitude $A_* = 3.1_{-0.9}^{+1.3}\times 10^{-15}$ at 68\% credibility for the spectral index $\gamma = 3.90\pm 0.40$ using data spanning 18 years~\cite{Zic:2023gta, Reardon:2023zen, Reardon:2023gzh}. EPTA/InPTA reports the amplitude $\log_{10}A_* = -14.54_{-0.41}^{+0.28}$ with a credibility of 90\% and with the spectral index $\gamma = 4.19_{-0.63}^{+0.73}$, using the full set of data spanning 24.7 years~\cite{EPTA:2023sfo, EPTA:2023akd, EPTA:2023fyk, EPTA:2023gyr, EPTA:2023xxk, EuropeanPulsarTimingArray:2023egv}. CPTA has reported a correlated signal with the amplitude $\log_{10}A_* = -14.4_{-2.8}^{+1.0}$ for a spectral index in the range $\gamma = [0.0, 6.6]$~\cite{Xu:2023wog}. 

\begin{figure}
	\includegraphics[width = \linewidth]{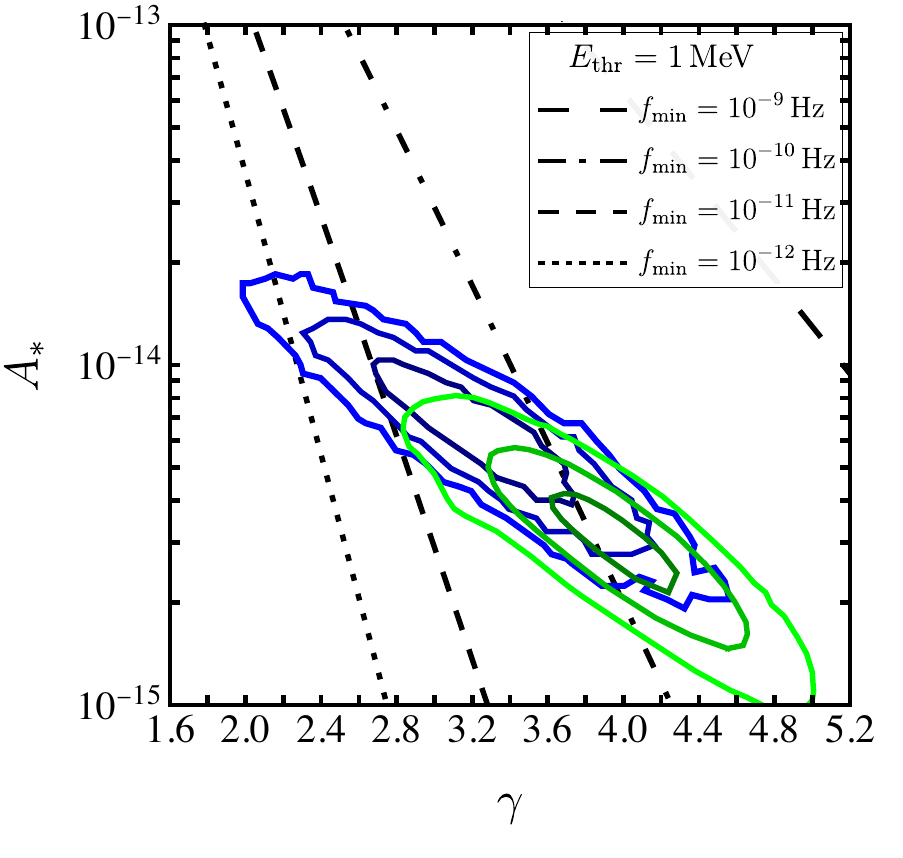} 
	\caption{The black contour lines bound the region in the parameter space $(\gamma, A_*)$ above which flavor oscillations in a neutrino wave packer are suppressed for different values for the minimum frequency $f_{\rm min}$. The threshold energy of the detector is fixed at $E_{\rm thr}=1\,\mathrm{MeV}$. Results are compared with the fit obtained by the NANOGrav collaboration~\cite{NANOGrav:2023gor} after analyzing the 15 years data set (blue) and the PPTA collaboration~\cite{Reardon:2023gzh} (green), with different shades marking the 1, 2 and 3$\sigma$ confidence regions.}
	\label{fig:EthrGWs}
\end{figure}

Fig.~\ref{fig:EthrGWs} shows the relation between $A_*$ and $\gamma$ for different choices of $f_{\rm min} \in [10^{-3}-1]$\,nHz. We assume that the energy threshold below which neutrino oscillations from a galactic source are suppressed is $E_{\rm thr} = 1\,$MeV, which is of the same order as the threshold energy that can be observed in near-future neutrino detectors. In fact, various techniques are being employed for the detection in present and near-future facilities. In the Jiangmen Underground Neutrino Observatory (JUNO)~\cite{JUNO:2015zny}, one of the detection channels occurs via the inverse beta decay (IBD) interactions of electron antineutrinos off the protons in the water Cherenkov tank, $\bar\nu_e + p \to e^+ + n$, with the reaction threshold $E_{\rm thr}^{\rm IBD} = 1.8\,$MeV; even lower threshold energies of $\mathcal{O}(0.2)\,$MeV are expected to be probed via the elastic neutrino-proton scattering channel $\nu_i + p +\to \nu_i + p$ for a neutrino species $i$~\cite{JUNO:2021vlw}. Hyper-Kamiokande~\cite{Hyper-Kamiokande:2018ofw} is also designed to operate via IBD and detect antineutrinos of energies close to $E_{\rm thr}^{\rm IBD}$, thanks to the recent introduction of the gadolinium doping~\cite{Super-Kamiokande:2008mmn, Super-Kamiokande:2021the} that allows for the discrimination of SNe neutrinos from the antineutrino background from nuclear power reactors. In the Deep Underground Neutrino Experiment (DUNE)~\cite{DUNE:2020ypp, DUNE:2020lwj}, electron neutrinos are mainly detected through their charge-current interactions with argon ($\nu_e + {}^{40}{\rm Ar} \to e^- + {}^{40}{\rm Kr}^*$) and elastic-scattering channels of neutrinos off charged leptons, with the experimental threshold $E_{\rm thr}^{\rm DUNE} \approx 5\,$MeV~\cite{Capozzi:2018dat}.

The initial composition is a major uncertainty in the analysis, as different astrophysical sources lead to an asymmetric flavor emissions of electrons and muon neutrinos~\cite{Lipari:2007su, Bustamante:2015waa}. Oscillation in a matter envelope can further complicate the picture by allowing for the appearance of $\nu_\tau$~\cite{Dighe:1999bi, Lunardini:2000fy, Razzaque:2009kq}. We have assumed an initial composition is predominant in electronic neutrinos, although similar results can be drawn if any of the other two flavors dominates the composition at the source. A flavor mixing at the source would make the discerning of the oscillations pattern more challenging with respect to the case of an initial beam of electronic neutrinos; however, the threshold energy related to the coherence term would not be altered by the initial composition.

We also report the results from the NANOGrav 15-year~\cite{NANOGrav:2023gor} (blue) and the PPTA~\cite{Reardon:2023gzh} (green) collaborations for 1$\sigma$, 2$\sigma$ and 3$\sigma$ confidence levels. If the spectrum measured by the PTA analyses holds down to the cutoff spectrum $f_{\rm min} \lesssim 10^{-10}\,$Hz, the suppression in the neutrino flavor oscillations below the energy threshold would be detectable by near-future neutrino detectors.
We have reported the results for frequencies well below the result in Eq.~\eqref{eq:fmin} for three reasons: i) there are uncertainties in the actual value of the mechanisms leading to the value for $T_{\rm insp}$ which could lower the minimum frequency further; ii) there could exist populations of ultra-massive BHs~\cite{Natarajan:2008ks, Carr:2020erq}; iii) the leading physical phenomenon observed by the PTA measurements is some exotic release of GWs with a strain extending to lower values of $f_{\rm min}$. If the minimum frequency is in the range of PTA methods, the phenomenon discussed will be detectable in near-future neutrino facilities. On the other hand, there would be a way to infer the cut-off frequency related to $A_*$ and $\gamma$ from future observations.

\begin{figure}
	\includegraphics[width = \linewidth]{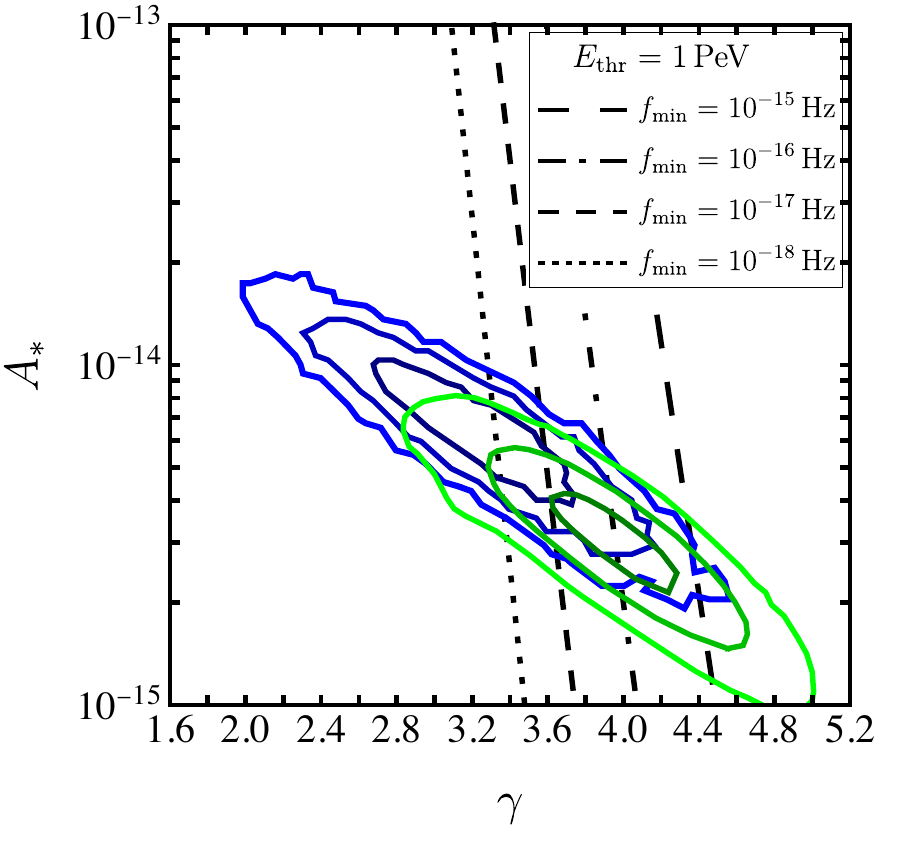} 
	\caption{Same as Fig.~\ref{fig:EthrGWs} for a minimum frequency in the ELF band, $f_{\rm min} = [10^{-18}, 10^{-15}]\,$Hz. The threshold energy of the detector is fixed at $E_{\rm thr}=1\,\mathrm{PeV}$.}
	\label{fig:EthrGWsPeV}
\end{figure}

We repeat the analysis by considering a lower cutoff frequency in the ELF band and the GWB spectrum measured by the NANOGrav and the PPTA collaborations. It results that the phenomenon affects neutrinos in the PeV energy range. Setting the threshold energy in Eq.~\eqref{eq:thresholdE} as $E_{\rm thr} = 1\,$PeV, the PTA results are consistent with a wide range of frequencies $f_{\rm min} = [10^{-18}, 10^{-15}]\,$Hz as shown in Fig.~\ref{fig:EthrGWsPeV}.

\section{Results for the SMBHBs fiducial model}

We now turn to the assessment of neutrino oscillations for the case where the pulsar timing results from the collective superposition of the GWB from incoherent SMBHB sources. For this model, characterized by a GWB strain $h_c(f) \propto f^{-2/3}$~\cite{Wyithe:2002ep} and a corresponding spectral index $\gamma = 13/3$, the NANOGrav collaboration reports the amplitude $A_* = 2.4_{-0.6}^{+0.7}\times 10^{-15}$ with 90\% credibility at the reference scale $f_{\rm yr}$~\cite{NANOGrav:2023gor}. Consistent recent results include $A_* = (2.5\pm 0.7)\times10^{-15}$ from the analysis by the EPTA/InPTA collaboration~\cite{EPTA:2023fyk} and $A_*=2.04_{-0.22}^{+0.25}\times10^{-15}$ as reported by the PPTA consortium~\cite{Reardon:2023gzh}.

\begin{figure}
	\includegraphics[width = \linewidth]{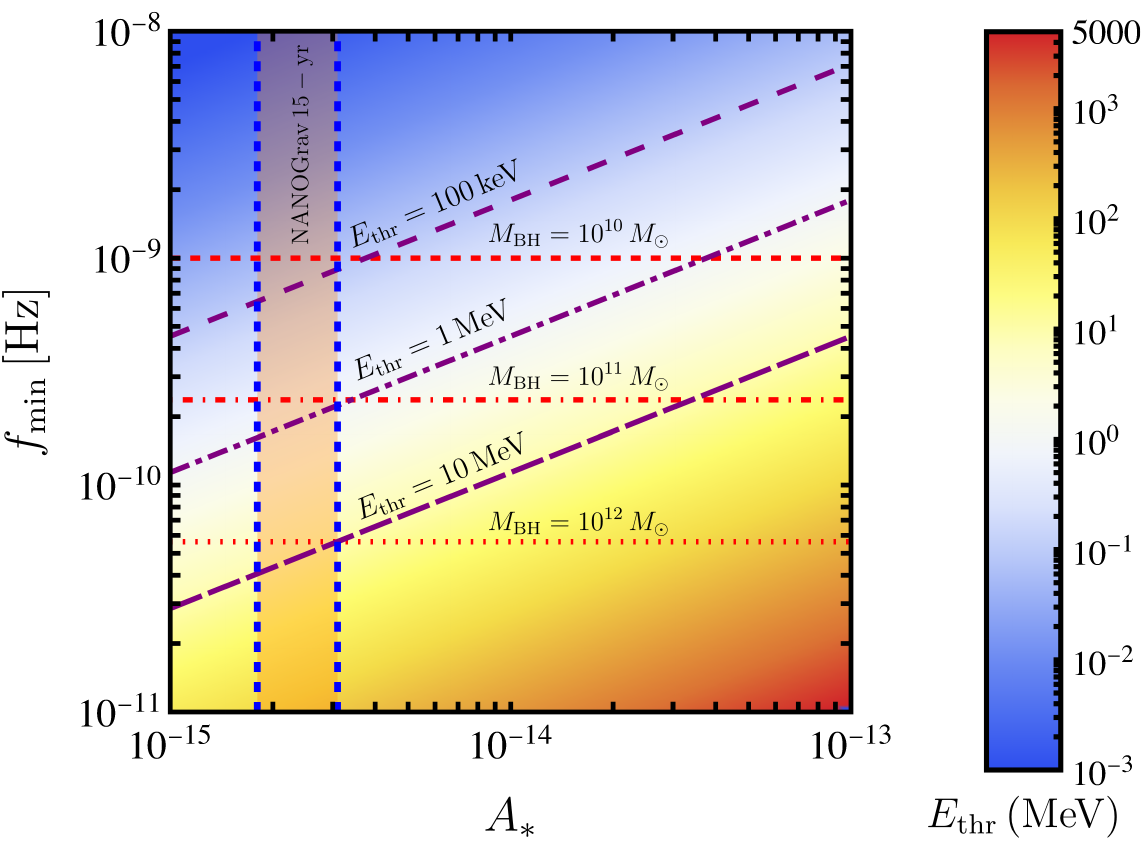} 
	\caption{The threshold energy $E_{\rm thr}$ in MeV below which flavor oscillations in the neutrino propagation are suppressed, for an outburst of neutrinos detected from a nearby supernova for $\gamma=13/3$. The vertical axis shows the minimal frequency at which Eq.~\eqref{eq:GWMBHB} holds and the horizontal axis shows $A_*$ value. Also shown are the black hole mass $M_{\rm BH}$ yielding $f_{\rm min}$ (horizontal red dashed line) and the threshold energy (purple dotted-dashed line).}
	\label{fig:AsvsD}
\end{figure}

Fig.~\ref{fig:AsvsD} shows the threshold energy $E_{\rm thr}$ as a function of $A_*$ and $f_{\rm min}$ for the spectral index $\gamma = 13/3$. The vertical band labeled ``NANOGrav 15-yr'' marks the amplitude of the GW strain reported in Ref.~\cite{NANOGrav:2023gor} at the reference scale $f_{\rm yr}$. The horizontal red lines show the value of $f_{\rm min}$ from Eq.~\eqref{eq:fmin} assuming a cutoff in the SMBHB distribution as $M_{\rm BH} = 10^{10}\,M_\odot$ (dashed line), $M_{\rm BH} = 10^{11}\,M_\odot$ (dot-dashed line), and $M_{\rm BH} = 10^{12}\,M_\odot$ (dotted line). In the region favored by the PTA analyses, the SMBHB populations of masses $M_{\rm BH} \gtrsim \mathcal{O}(10^{12})M_\odot$ lie below the threshold $E_{\rm thr} = 10\,$MeV (purple dashed line), while frequencies below $f_{\rm min} \approx \mathcal{O}(10^{-10}{\rm \,Hz})$ are accessible if the threshold energy is pushed to the value $E_{\rm thr} = 1\,$MeV considered when discussing Fig.~\ref{fig:EthrGWs} (purple dot-dashed line). The detector energy threshold in a large liquid-scintillator detector such as JUNO would be limited by the intrinsic radioactive backgrounds of the experiment itself and it could then be significantly reduced by controlling the abundance of ${}^{14}$C as already achieved in BOREXINO~\cite{Borexino:2000uvj}. Under these conditions, the energy threshold is as low as $E_{\rm thr}^{\rm JUNO} = 100\,$keV (purple finely dashed line) and allows for the discrimination of the energy at which the suppression in the neutrino flavor oscillations occurs up to the nHz range.

\section{Discussion}
\label{sec:Discussion}

The effects of the GWB on neutrino oscillations in the vacuum would not be easily observed at Earth due to the phenomenon of decoherence~\cite{Kiers:1995zj, Giunti:1997sk}. Indeed, considering a neutrino wave packet emitted from a distant supernova, characterized by a sizeable energy spread $\sigma_E \sim E_\nu$ that corresponds to an initial size of the packet $\sigma_x \sim 10^{-13}\,$m, it results in a coherence length $L_{\rm coh} = 2E_\nu^2\sigma_x/\Delta m_{12}^2$. Since for $E_\nu<\mathcal{O}(\mathrm{TeV})$ $L_{\rm coh}$ is much smaller than astronomical distances, neutrino decoherence from distant sources occurs rather quickly, leading to an incoherent ensemble of mass eigenstates, with coherence effects that cannot be picked up in present detectors~\cite{Hansen:2016klk, Akhmedov:2017mcc}. \\
Moreover, let us consider a detector on Earth with the energy resolution smaller than the wave packet spread, $\delta E < \sigma_E$, so that the detection of neutrino oscillations which are coherent over the distance $D$ requires the energy resolution
\begin{equation}
    \epsilon \equiv \frac{\delta E}{E_\nu} < \frac{2 E_\nu}{\Delta m_{12}^2}\frac{1}{D}\,.
\end{equation}

Therefore, in order to detect oscillations from coherent neutrinos over galactic scales, we consider the case of very energetic neutrinos discussed in Fig.~\ref{fig:EthrGWsPeV}, for which the energy threshold corresponds to $E_{\rm thr} = (10^{-2} - 1)\,$PeV and the coherence length scale is $\sim $kpc for energies $E_\nu = 100\,$PeV and resolution $\epsilon = 0.1\%$. Various experiments targeting very energetic neutrinos are currently under construction with different methodologies that involves the deployment on ice or water, including the IceCube Neutrino Observatory~\cite{IceCube-Gen2:2020qha}, the proposed detectors ORCA and ARCA as part of the KM3NeT collaboration~\cite{KM3Net:2016zxf}, the Pacific Ocean Neutrino Experiment (P-ONE)~\cite{P-ONE:2020ljt, Resconi:2021ezb}, and the tRopIcal DEep-sea Neutrino Telescope (TRIDENT)~\cite{Ye:2022vbk}.

Neutrinos that are less energetic might still exhibit oscillations at detection as coherence in neutrino wave packets might re-emerge once neutrinos propagate inside the Earth layers in the so-called ``catch-up'' effect~\cite{Kersten:2015kio}. Indeed, the latter can compensate for the wave packets separation, leading to the possibility of observing neutrino oscillations. The GWB interaction prior the onset of the catch-up effect could lead to a reduced coherent length and the lack of oscillations only below a certain energy threshold.

The propagation of astrophysical neutrinos might also help study more exotic scenarios in which the nature of neutrinos is that of a pseudo-Dirac particle~\cite{Wolfenstein:1981kw, Petcov:1982ya, Bilenky:1983wt}, for which a tiny mass split $\delta m^2$ between the particle and its antiparticle exists in the mass eigenstates. Due to the small mass split, oscillations between active and sterile neutrino states are naturally coherent over astronomical scales~\cite{DeGouvea:2020ang}, so that the effect we underline in this work would impact the propagation of pseudo-Dirac neutrino scenarios from SN1987A studied in Ref.~\cite{Martinez-Soler:2021unz}.

\section{Conclusions}

We have studied the effects of a gravitational wave background (GWB) on the propagation of astrophysical neutrinos and the detection of the pattern of oscillations at terrestrial facilities. Neutrinos produced by galactic sources would interact with the GWB while traveling to Earth and might be subject to the decoherence effect which suppresses their flavor oscillation. We assume that the GWB extends to frequencies above $f_{\rm min}$, below which other mechanisms are responsible for the energy loss of the system besides the release of GWs; moreover, no other source of GWs is present below this frequency that would significantly alter the integral expression in Eq.~\eqref{eq:decoherenceterm}. For example, the lowest frequency at which the GWB from inflation is expected corresponds to the size of the comoving horizon when the Big Bang Nucleosynthesis occurs and it is $\mathcal{O}(0.1{\rm\,nHz})$~\cite{Vagnozzi:2020gtf, Benetti:2021uea, Vagnozzi:2022qmc}. Modelling the GWB strain in terms of the amplitude $A_*$ and an index $\gamma$ as in Eq.~\eqref{eq:strain} leads to the expected coherence term $\Gamma^{(ij)}$ that modulates the flavor conversion.

The GWB is modelled according to the detection recently reported by the PTA collaborations upon independent analyses of the data sets collected over a multi-decadal time frame. We have considered separately the results obtained from an unconstrained fit of the power-law in Eq.~\eqref{eq:strain} with variable amplitude and spectral index $(A_*, \gamma)$, as well as a fiducial model with a fixed spectral index $\gamma = 13/3$. This results in two main implications:

i) For the variable power-law exponent model, the suppression in neutrino oscillations could be detectable if indeed $f_{\rm min} \lesssim 10^{-10}\,$Hz. In this scenario, the neutrino energy below which the flavor suppression occurs could be inferred from the corresponding neutrino flux detected in near-future neutrino facility. Fig.~\ref{fig:EthrGWs} compares the results from NANOGrav 15-year and the PPTA data analyses over the $(A_*,\gamma)$ parameter space with the reach at neutrino facilities assuming the energy threshold in detection $E_{\rm thr} \simeq 1\,$MeV and different values of $f_{\rm min}$ in the model.
 
ii) For a fixed spectral index $\gamma = 13/3$, the region $A_* \approx 2\times 10^{-15}$ inferred by the PTA data analyses in Fig.~\ref{fig:AsvsD} is within experimental reach if $f_{\rm min} \lesssim 10^{-10}\,$Hz for $E_{\rm thr} \simeq 1\,$MeV, see Fig.~\ref{fig:AsvsD}, or even $f_{\rm min} \lesssim 10^{-9}\,$Hz for $E_{\rm thr} \simeq 100\,$keV provided that the neutrino flux is sufficiently high to overcome the solar background at such low energies. This would allow to indirectly probe the GWB at frequencies below the nHz that are not detectable by using PTA methods.

The same considerations can be obtained referring to a $f_{\rm min} \lesssim 10^{-18}\,$Hz which lead to an $E_{\rm thr} \sim \mathcal{O}(1)\,$PeV as well discussed in Sec.~\ref{sec:Discussion}.

Future observations in neutrino facilities would improve their spatial and temporal resolutions as well as enable directional detection to correctly pin down the time window of the source and the oscillation pattern in the detector. As expressed above, indirect information on the GWB properties can be gathered by identifying a suppression in neutrino flavor oscillations. We have assumed that the initial composition of the flux is in electronic neutrinos; while this choice maximizes the effects discussed, some mixing could occur at production.

In conclusion, the pattern of flavor oscillations in a neutrino flux can be used to probe the GWB and test different astrophysical models. The role of near-future detectors will be crucial to detect the threshold at which the suppression of the neutrino flavor oscillations occurs and infer the properties of the GWB below the nHz.

\begin{acknowledgments}
We thank Orlando Perez for the constructive comments over an earlier version of the draft.
L.V.\ would like to thank Iwan Blake and Donglian Xu for an extensive discussion on neutrino detector strategies. The work of G.L.\ and L.M.\ is supported by the Italian Istituto Nazionale di Fisica Nucleare (INFN) through the ``QGSKY'' project and by Ministero dell'Istruzione, Universit\`a e Ricerca (MIUR). L.V.\ acknowledges support by the NSFC through the grant No.\ 12350610240, as well as hospitality by the INFN Frascati National Laboratories during the completion of this work. This publication is based upon work from the COST Actions ``COSMIC WISPers'' (CA21106) and ``Addressing observational tensions in cosmology with systematics and fundamental physics (CosmoVerse)'' (CA21136), both supported by COST (European Cooperation in Science and Technology).
\end{acknowledgments}

\bibliography{sources.bib}

\end{document}